\documentstyle[aps,12pt,amssymb,epsfig]{revtex}

\begin{document}

\baselineskip=0.75cm
\newcommand{\ini}{\begin{equation}}
\newcommand{\fin}{\end{equation}}
\newcommand{\inir}{\begin{eqnarray}}
\newcommand{\finr}{\end{eqnarray}}
\newcommand{\inif}{\begin{figure}}
\newcommand{\finf}{\end{figure}}
\newcommand{\bc}{\begin{center}}
\newcommand{\ec}{\end{center}}

\def\ol{\overline}
\def\pa{\partial}
\def\ra{\rightarrow}
\def\ts{\times}
\def\df{\dotfill}
\def\bs{\backslash}
\def\dg{\dagger}

$~$

\hfill DSF-21/2001

\vspace{1 cm}

\centerline{{\bf BARYOGENESIS VIA LEPTOGENESIS IN SO(10) MODELS}}

\vspace{1 cm}

\centerline{\large{F. Buccella, D. Falcone, F. Tramontano}}

\vspace{1 cm}

\centerline{Dipartimento di Scienze Fisiche, Universit\`a di Napoli,}
\centerline{Complesso di Monte S. Angelo, Via Cintia, Napoli, Italy}

\vspace{1 cm}

\begin{abstract}

\noindent
We discuss the baryogenesis via leptogenesis mechanism within the
supersymmetric and nonsupersymmetric SO(10) models. We find that the
nonsupersymmetric model, endowed with an intermediate scale,
is generally favoured, unless some fine tuning occurs in the supersymmetric
case.

\end{abstract}

\newpage

\section{Introduction}

The origin of the baryon asymmetry in the universe (baryogenesis) is a much
discussed topic \cite{rt}. A popular mechanism is the baryogenesis via
leptogenesis \cite{fy}, where the out-of-equilibrium decays of heavy
right-handed
Majorana neutrinos generate a lepton asymmetry which is partially transformed
into a baryon asymmetry by electroweak sphaleron processes \cite{krs}. A minimal
framework required is the standard model with heavy right-handed neutrinos,
but the mechanism is active also within unified theories \cite{pil}, and in
particular the $SO(10)$ model \cite{gfm},
which naturally contains heavy right-handed neutrinos.
The light left-handed Majorana neutrinos are obtained by means of
the seesaw mechanism \cite{ss}.

The baryogenesis via leptogenesis has been studied in many papers
\cite{bpp,cfl,bbb,ft1,ft2,jpr}.
In this letter we address a specific issue, not
explicitly considered before, namely the possibility to generate the baryon
asymmetry within the nonsupersymmetric $SO(10)$ model, characterized by the
presence of an intermediate mass scale where both lepton number conservation
and quark-lepton symmetry are broken. We match the result to the supersymmetric
case, where the two effects occur at the unification scale, and no intermediate
scale is present.

In section II we summarize the relevant formulas of the baryogenesis via
leptogenesis mechanism. In section III we calculate the baryon asymmetry
within the $SO(10)$ model by using two distinct forms for the mass matrices
of the right-handed neutrino, which correspond to the supersymmetric and
nonsupersymmetric cases, respectively. In section IV we give our conclusions.

\section{The baryogenesis via leptogenesis mechanism}

As proposed in Ref.\cite{fy}, a baryon asymmetry can be generated from a
lepton asymmetry. We define the baryon asymmetry as \cite{kt1}
\ini
Y_B =\frac{n_B-n_{\ol{B}}}{s}=
\frac{n_B-n_{\ol{B}}}{7 n_{\gamma}}=\frac{\eta}{7},
\fin
where $n_{B,\ol{B},\gamma}$ are number densities, $s$ is the entropy density
and $\eta$ is the baryon-to-photon ratio. The range of $Y_B$ required for
a successful description of nucleosynthesis is $Y_B=10^{-11}-10^{-10}$, see
for example Ref.\cite{osw}. In the baryogenesis via leptogenesis framework,
the baryon asymmetry is related to the lepton asymmetry \cite{htks},
\ini
Y_B=\frac{a}{a-1} Y_L,~~
a=\frac{8N_f+4N_H}{22N_f+13 N_H},
\fin
where $N_f$ is the number of families and $N_H$ the number of light Higgs
doublets. For $N_f=3$ and $N_H=1$ or 2 (standard or supersymmetric case),
it is $a \simeq 1/3$ and $Y_B \simeq -Y_L/2$.

The lepton asymmetry is written as \cite{lv}
\ini
Y_L=d~ \frac{\epsilon_1}{g^*}
\fin
where $\epsilon_1$ is a CP-violating asymmetry produced by the decay of the
lightest heavy neutrino, $d$ is a dilution factor which takes into account the
washout effect of inverse decay and lepton number violating scattering, and
$g^*=106.75$ in the standard case or 228.75 in the supersymmetric case is the
number of light degrees of freedom in the theory.

In the standard case $\epsilon_1$ is given by \cite{crv}
\ini
\epsilon_1=\frac{1}{8 \pi v^2 (M_D^{\dg} M_D)_{11}}\sum_{j=2,3}
$Im$ [(M_D^{\dg} M_D)_{j1}]^2 f \left( \frac{M_j^2}{M_1^2} \right),
\fin
where $M_D$ is the Dirac neutrino mass matrix when
the Majorana neutrino mass matrix $M_R$ is
diagonalized with eigenvalues $M_i$ $(i=1,2,3)$, $v=175$ GeV is the VEV of
the standard model Higgs doublet, and
$$
f(x)=\sqrt{x} \left[1-(1+x) \ln \frac{1+x}{x}-\frac{1}{x-1} \right].
$$
In the supersymmetric case $v \ra v \sin \beta$,
$$
f(x)=-\sqrt{x} \left[ \ln \frac{1+x}{x} +\frac{2}{x-1} \right],
$$
and a factor $4$ is included in $\epsilon_1$, due to more decay
channels. For a hierarchical spectrum of heavy neutrinos we have
$f \sim M_1/M_j$. The formula (4)
is obtained by calculating the interference between the tree level and one loop
decay amplitudes of the lightest heavy neutrino, and includes vertex \cite{fy}
and self-energy \cite{ls} corrections.
The latter are dominant if $M_1$ and $M_j$ are nearly equal, in which case
an enhancement of the asymmetry may occur.

The dilution factor should be obtained by solving the Boltzmann equations.
We use an approximate solution \cite{kt2,pil,fp,ft1}:
\ini
d=\frac{0.24}{k(\ln k)^{0.6}}
\fin
for $k \gtrsim 10$, and
\ini
d=\frac{1}{2k},~~d=1
\fin
for $1 \lesssim k \lesssim 10$, $0 \lesssim k \lesssim 1$, respectively,
where the parameter $k$ is
\ini
k = \frac{M_P}{1.7 v^2 32 \pi \sqrt{g^*}}\frac{(M_D^{\dg} M_D)_{11}}{M_1},
\fin
and $M_P$ is the Planck mass.

The baryon asymmetry depends on both the Dirac and the Majorana mass
matrices of neutrinos. In the following section we adopt general
approximate forms for these matrices and study the implications for
leptogenesis.

\section{Leptogenesis in SO(10) models}

In unified $SO(10)$ models, $M_R$ is generated from the Yukawa coupling of
right-handed neutrinos with the Higgs field that breaks the unification or the
intermediate symmetry down to the standard model, see for example
Ref.\cite{dk}. When such a Higgs field
takes a VEV, the right-handed neutrinos get a Majorana mass.
This happens because lepton number is broken at that scale. Therefore, in
the supersymmetric
case the mass scale of the right-handed neutrino is similar to the unification
scale, $M_R \sim M_U \sim 10^{16}$ GeV, while in the nonsupersymmetric case
the scale of $M_R$ is about the intermediate scale, $M_R \sim M_I \sim 10^{11}$
GeV \cite{lmpr}.
On the other hand, $M_U$ or $M_I$ are also the scale of the quark-lepton
symmetry, that is the gauge subgroup $SU(4) \ts SU(2)_L \ts SU(2)_R$, where
the $SU(4)$ component includes the lepton number as fourth color \cite{ps}.
This framework gives $M_d \sim M_e$ and $M_u \sim M_{\nu}$, where $M_{d,u}$
are quark mass matrices, $M_e$ is the charged lepton mass matrix and $M_{\nu}$
is the Dirac neutrino mass matrix. The light (effective) neutrino mass matrix
$M_L$ is obtained by means of the seesaw formula
\ini
M_L = - M_{\nu} M_R^{-1} M_{\nu}.
\fin
Since quark mixing is small, that is quark mass matrices are nearly diagonal,
from quark-lepton symmetry we get nearly diagonal $M_e$ and $M_{\nu}$.

By inverting formula (8) with respect to $M_R$, in Ref.\cite{f1} the
approximate structures leading to the unification and intermediate scales were
identified. The matrix $M_R$ should be nearly diagonal or nearly offdiagonal,
respectively, with a strong hierarchy in the former case and a more moderate
hierarchy in the latter. These two situations are similar to those discussed
in Ref.\cite{smir} in order to get a seesaw enhancement of lepton mixing.
The condition $M_{R33} \simeq 0$ has been
further discussed in Ref.\cite{ab}(see also Ref.\cite{js}). We 
assume large mixing of solar and atmospheric neutrinos.
For the Dirac neutrino mass matrix we take
\ini
M_{\nu} = \frac{m_{\tau}}{m_b}~ \text{diag}(m_u,m_c,m_t),
\fin
where the ratio $m_{\tau}/m_b$ takes into account the running of quark masses
with respect to lepton masses. For the mass matrix of right-handed neutrinos
we take in the supersymmetric case \cite{f1}
\ini
M_R \simeq \left( \frac{m_{\tau}}{m_b} \right)^2
\left( \begin{array}{ccc}
m_u^2 & -\frac{m_u m_c}{\sqrt2} & \frac{m_u m_t}{\sqrt2} \\
-\frac{m_u m_c}{\sqrt2} & \frac{m_c^2}{2} & -\frac{m_c m_t}{2} \\
\frac{m_u m_t}{\sqrt2} & -\frac{m_c m_t}{2} & \frac{m_t^2}{2}
\end{array} \right) \frac{1}{2m_1},
\fin
where $m_1 \sim 10^{-3}$ eV is the mass of the lightest effective neutrino,
and in the nonsupersymmetric case \cite{f1,ab}
\ini
M_R \simeq \left( \frac{m_{\tau}}{m_b} \right)^2
\left( \begin{array}{ccc}
m_u^2 & -\frac{m_u m_c}{\sqrt2} & \frac{m_u m_t}{\sqrt2} \\
-\frac{m_u m_c}{\sqrt2} & 2 U_{e3}{m_c^2} & x \\
\frac{m_u m_t}{\sqrt2} & x & 0
\end{array} \right) \frac{1}{m_1},
\fin
where $U_{e3} \simeq 0.1$ is the element 1-3 in the lepton mixing matrix and $M_{R23}$
takes values from $10^{10}$ to $10^{12}$ GeV. Note that $M_{R13} \sim
10^{11}$ GeV. We will discuss the different
implications for leptogenesis of matrices (10) and (11), which correspond to
distinct models. Since our main interest is the general result,
especially the difference between the supersymmetric and nonsupersymmetric
cases, we do not include phases and drop the imaginary part in Eqn.(4).
We diagonalize $M_R$ by the rotation $U_R^T M_R U_R$, so that $M_D=M_{\nu}U_R$,
and insert both $M_i$ and $M_D$ is Eqn.(4). In this way the baryon asymmetry
can be determined.

In the supersymmetric model we find $Y_B \sim 10^{-15}-10^{-14}$, where
the range corresponds to moderate changes in $M_R$. We use $\sin \beta \simeq 1$
for quark-lepton Yukawa unification of the third generation \cite{lp}, although the
baryon asymmetry depends very weakly on this parameter. The value of $k$ is
slightly larger than 1. This case is similar to the one studied in
Ref.\cite{ft1}, where a sufficient amount of asymmetry is obtained only
by fine tuning of some neutrino parameters.

In the nonsupersymmetric model, the baryon asymmetry is around the required
order of magnitude $Y_B \sim 10^{-11}$. In Fig. 1 we plot the result as a function
of Log$_{10} M_{R23}$. Here we have $k \sim 10-10^{3}$.
For lower values of $M_{R23}$ the baryon asymmetry
undergoes a moderate increase and for higher values it drops towards the
supersymmetric result. 

\newpage

\begin{figure}[ht]
\begin{center}
\epsfig{file=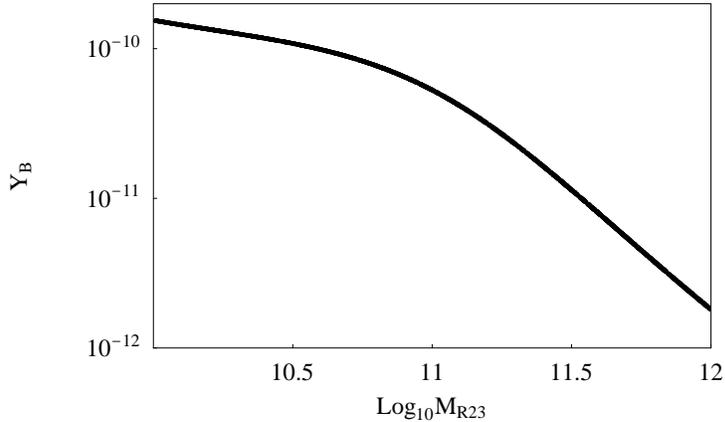,height=6cm}
\caption{The baryon asymmetry $Y_B$ vs. Log$_{10}M_{R23}$ in the model with
intermediate scale.}
\end{center}
\end{figure}

\section{Conclusion}

The main result of the present paper is that the nonsupersymmetric $SO(10)$
model is favoured for leptogenesis with respect to the supersymmetric model.
In fact, in the latter case a sufficient amount of baryon asymmetry can be
obtained only by means of fine tuning, while the nonsupersymmetric model
gives a baryon asymmetry of the same order as required.

By matching the present result with previous work \cite{ft1,ft2}, we realize
that the supersymmetric model with full quark-lepton symmetry generally gives
a too small asymmetry \cite{ft1}. This can be avoided within the $SU(5)$
model, where $M_{\nu}$ is no more related to $M_u$, by taking a moderate
hierarchy in $M_{\nu}$ \cite{ft2}, or in the nonsupersymmetric model by
means of a roughly offdiagonal $M_R$, corresponding to a moderate
hierarchy of its eigenvalues.

\end{document}